\documentclass[twocolumn,showpacs,prl]{revtex4}

\usepackage{graphicx}%
\usepackage{dcolumn}
\usepackage{amsmath}
\usepackage{latexsym}

\begin {document}

\title{Precise toppling balance, quenched disorder, and universality for sandpiles}
\author
{
R. Karmakar$^1$, S. S. Manna$^1$, and A. L. Stella$^2$
}

\affiliation
{
$^1$Satyendra Nath Bose National Centre for Basic Sciences Block-JD, Sector-III, Salt Lake, Kolkata-700098, India \\
$^2$INFM - Dipartimento di Fisica and Sezione INFN, Universit\`a di Padova, I-35131 Padova, Italy
}

\begin{abstract}

A single sandpile model with quenched random toppling matrices captures the crucial features
of different models of self-organized criticality.
With symmetric matrices avalanche statistics falls in the multiscaling
BTW universality class. In the asymmetric case the simple scaling of the Manna
model is observed. The presence or absence of a precise toppling balance between the amount 
of sand released by a toppling site and the total quantity the same site receives 
when all its neighbors topple once determines the appropriate universality class. 

\end{abstract}

\pacs{05.65.+b  
      05.70.Jk, 
      45.70.Ht  
      05.45.Df  
}
\maketitle

Self-organized criticality (SOC) occurs in the non-linear transport
of some physical entity, like energy, sand, or stress, through
a system of linear size $L$ under a constant, slow external drive \cite {Bakbook,Dhar1,Grass,Man}.
The transport has breakdown features, with intermittent bursts of
activity called avalanches. At long times the probability distribution
of avalanche sizes becomes critical since they do not reveal characteristic
scales between $L$ and the minimal length specified in the model. 
In spite of many efforts, the physical mechanisms underlying the different 
forms of scaling realized in such models and the related universality issues
remain poorly understood. The probability distribution of the number
of topplings in an avalanche of the Bak, Tang and Wiesenfeld (BTW) \cite {Bak} sandpile,
historically the prototype of SOC, has been shown recently to violate the 
finite size scaling (FSS) ansatz, assumed for many years, and to obey a peculiar 
form of multiscaling \cite {Stella1,Stella2}. On the other hand, in the Manna
stochastic sandpile \cite {Manna}
the corresponding distribution is known to obey FSS \cite {Lubeck,Stella2,Biham2}.
Understanding the key mechanisms at the basis
of these radically different forms of scaling remains a major challenge
which should be faced also in the perspective of new concrete 
applications \cite {Bersh}.

In this Letter we study sandpile models \cite {Dhar1} on lattices with quenched disorder.
This means that different bonds of the lattice allow flow
of different, but constant numbers of sand grains through them. 
We further distinguish between
an undirected case, in which the flow is identical in the two directions 
for each bond, and a directed one, in which this flow can be asymmetric.
We show that, once averaged over the possible 
realizations of the quenched disorder, the avalanche size
distribution of the model with symmetric flow in each bond obeys the same 
multiscaling as the
BTW model, while asymmetry in the flow leads to FSS with the
exponents of the Manna sandpile. Thus, for a disordered sandpile
the two main SOC universality classes
can both be realized simply upon enforcing or releasing a local symmetry 
requirement.

The deterministic sandpile model on a square lattice
of size $L$ is described using an integer `toppling matrix' (TM) $\Delta$ 
of size $L^2 \times L^2$ \cite {Dhar1}. At lattice site $k$ there is
a column of $h_k$ sand grains. The system is externally driven by 
adding a single grain at a time at a randomly selected site $i$: $h_i \to h_i+1$. 
Each site has a threshold height $H_i$ of stability. If 
$h_i > H_i$, the 
sand column at $i$ topples and grains are distributed to other sites.
Consequently, all heights are updated as: $h_j \rightarrow h_j - \Delta_{ij}$
where $\Delta_{ii}>0$ and $\Delta_{ij} \le 0$ for $i \neq j$.
The threshold heights are chosen as $H_i = \Delta_{ii}$.
Grain conservation is assured by putting $\Delta_{ii}=-\sum_{j \neq i} \Delta_{ij}$.
Grains must flow out of the system through the boundary sites to maintain
a stationary state. The BTW model is a special case of this formulation where
$\Delta_{ii}=4$, $\Delta_{ij}=-1$ 
for $|i-j|=1$, and $\Delta_{ij} = 0$ for $|i-j| > 1$ \cite {Bak}.
In contrast, in the stochastic sandpile model \cite {Manna},
annealed randomness enters in the grain distribution
process upon toppling. Indeed, in this case each grain of the toppling site is
transferred to a randomly selected neighboring site.
For both BTW and stochastic sandpile grain number is conserved, boundaries 
are open, and, when
a site topples grains are distributed to its neighborhood.
In spite of these common features
and similarities \cite{Shilo}, the difference in behavior of the two models concerns even the
type of scaling.

In the Manna sandpile, the randomness in the choice of the neighbors
getting a grain from the toppling site can be regarded as
an ``annealed'' disorder.
Indeed, the random choice of the TM elements is constantly
updated during the course of a given avalanche. In contrast,
a ``quenched'' disorder is realized in the models discussed here,
for which a random realization of $\Delta$ 
is maintained for full samples of avalanches, and an average over 
the distributions of many samples created is performed eventually. 

Our $\Delta_{ij}$ is nonzero only for $i=j$ or $|i-j|=1$. However,
unlike in the BTW model, $\Delta_{ij}$ takes random values.
These are chosen among the negative integers 
$-1, -2, .., -m$ when $|i-j|=1$. 
Thus, when one of the sites
joined by a nearest neighbor bond topples, this bond can allow flow
of more than one grain, unlike in the BTW case.
When the TM is symmetric we call the model undirected and assign 
only one random integer
$\Delta_{ij}=\Delta_{ji}$ independently to each nearest neighbor bond (Fig. 1(a)). 
In the directed case the TM is asymmetric and both $\Delta_{ij}$ and $\Delta_{ji}$ are assigned
independently drawn random integers (Fig.1(b)).
In this way a toppling at one end of a 
bond may send, through the same bond, a different number of grains from that at the other end.
At any site $i$ we put $H_i = \Delta_{ii} = -\sum_{j\neq i} \Delta_{ij}$. 
The total number of grains received
by the site $i$ when all neighboring sites topple once is
$H'_i = -\Sigma_{j\neq i} \Delta_{ji}$. 
We call $H_i$ and $H'_i$ the out-degree and in-degree of site $i$,
respectively.

\begin{figure}[top]
\begin{center}
\includegraphics[width=6.0cm]{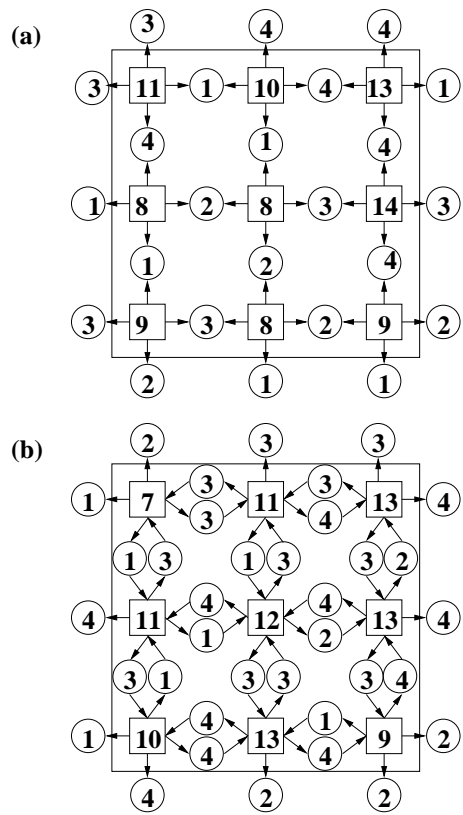}
\end{center}
\caption{
    Random grain flow distribution along the bonds of a
  $3 \times 3$ square lattice: (a) undirected case,
  (b) directed case.
}
\label{Fig1}
\end{figure}

Irrespective of whether $\Delta$ is symmetric or not, the models
enjoy the Abelian property of the BTW sandpile \cite {Dhar2}. This allows 
to establish a number of general exact results
concerning the stationary state, recurrent configurations, etc.,
which hold in both the directed and the undirected case \cite {Dhar1}.
An important notion is that an avalanche can be decomposed into
waves of toppling \cite {Prie1,Prie2}.
To this purpose one considers the sequence of topplings following
the addition of the seed grain at site $0$. The first wave is the set of
all topplings that follows the first toppling at $0$ while $0$ itself
is prevented from a possible second toppling.
If $0$ is still unstable after the first wave,
the second wave starts by allowing a second toppling at $0$, 
and so on.

\begin{figure}[top]
\begin{center}
\includegraphics[width=5.5cm]{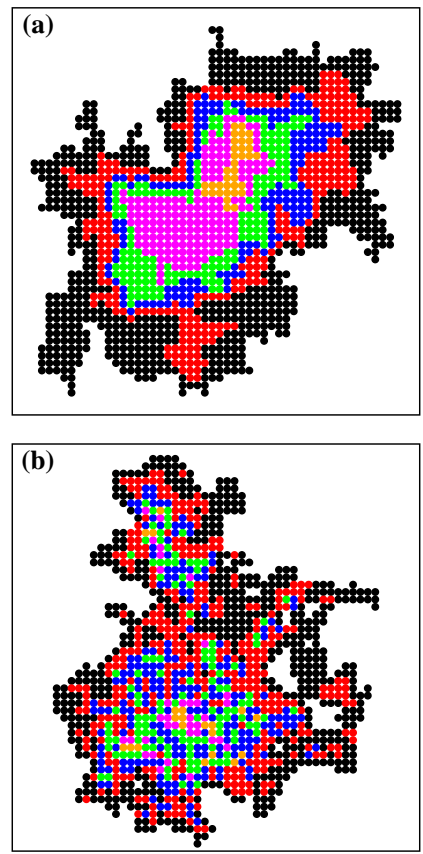}
\end{center}
\caption{
Multiply toppled sites within avalanches
are shown by circles of different colors:
1 (black), 2 (red), 3 (blue),
4 (green), 5 (magenta), 6 (orange) for the
(a) undirected model (b) directed model.
}
\label{Fig2}
\end{figure}

A peculiar property of waves 
with a symmetric TM is that the set of lattice sites which topple
has no holes. For example, a single untoppled
site can not be fully surrounded by toppled sites. Indeed, in 
the undirected model the equality $H_i=H'_i$ is
strictly maintained at all sites except at the boundary, which implies that a site must topple 
irrespective of its height, if all its neighbors topple once. Thus,
waves in the undirected model, like in the BTW model, 
have no holes.
In addition one can show that all sites involved by a wave 
topple only once, like the seed site. 
All this is not true anymore with asymmetric TM.
In this case the waves can have holes and can include sites which topple 
more than once. 
Indeed, in the directed model, $H_i \ne H'_i$ in general
and a site with $H_i > H'_i$
does not topple even if all its neighbors topple once. This creates
a single site hole in the avalanche. On the other hand, if a site $i$ has
out-degree sufficiently smaller than its in-degree, it may topple for the second time
even if none of its neighbors have toppled for the second time. Thus, in general
different sites belonging to a given wave topple a different number of times.

 The compactness and uniformity of waves in the undirected model
leads us to expect for them and for the resulting avalanches a structure
similar to that of the BTW model (Fig. 2(a)). The non-uniformity introduced by 
disorder should not be relevant at large scales when one counts toppling
events. The situation is quite different for the directed model: the structures
of waves and of avalanches in this case (Fig. 2(b)) look in fact similar to those
of the Manna model \cite {Stella3} where the numbers of grains transmitted upon in the two directions of a 
given bond are different in general, and this imbalance is maintained dynamically 
due to the stochastic distribution of sand grains.

\begin{figure}[top]
\begin{center}
\includegraphics[width=7.0cm]{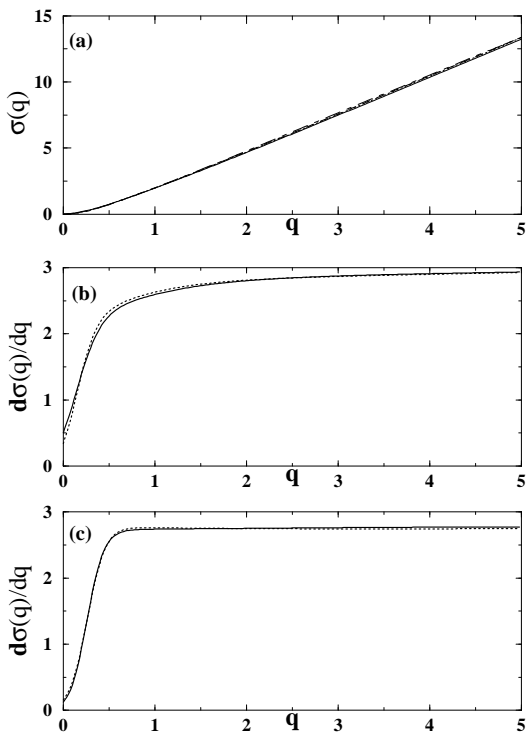}
\end{center}
\caption{
(a) Plot of  $\sigma (q)$ vs. $q$ for the BTW (solid), undirected (dotted), Manna (dashed), directed (dot-dashed) models.
Comparison of $d\sigma (q)/dq$ vs. $q$ plots between
(b) BTW (solid lines) and undirected (dotted lines) models and for the
(c) Manna (solid lines) and directed (dotted lines) models.
}
\label{Fig3}
\end{figure}

In the BTW model, the probability distribution ${\rm Prob}(s,L)$ of the total number
of topplings, $s$, in an avalanche has been found recently
to obey a multiscaling ansatz~\cite {Stella2}.
On the other hand, it is pretty well established \cite {Lubeck}
by now that in the Manna stochastic sandpile this distribution
obeys simple FSS as:
\begin{equation}
{\rm Prob}(s,L) \sim s^{-\tau} f(\frac {s}{L^{D}} ), \quad
\end{equation}
where the scaling function $f(x) \sim constant $ in the limit of $x \to 0$ and
$f(x)$ approaches zero very fast for $x >> 1$. The exponent $\tau$ and 
the dimension $D$ fully characterize the scaling of ${\rm Prob}$ in this case.
One immediate way to check validity of Eqn. (1) is to attempt a data collapse
by plotting $s^{\tau} {\rm Prob}$ vs. $s/L^D$ with trial values of the exponents.
We collected extensive data for both our models ($m=4$) for $L$ = 128, 256, 512, 1024, 2048 and 4096,
namely 50 million avalanches in 500 independent configurations for $L=128$ down to
$\approx$ 1.1 million avalanches for 9 configurations for $L$ = 4096.
The first $\sim 4L^2$ avalanches were skipped to reach the steady state.
For the directed case collapse works very nicely giving $\tau \approx 1.28$
and $D \approx 2.75$, close to the most reliable estimates of
the Manna sandpile exponents~\cite {Lubeck}. For the undirected model the
collapse does not work for a single set of $\tau$ and $D$ and for all values of $s$ and $L$.
This is found also in the BTW sandpile.

\begin{figure}[top]
\begin{center}
\includegraphics[width=6.0cm]{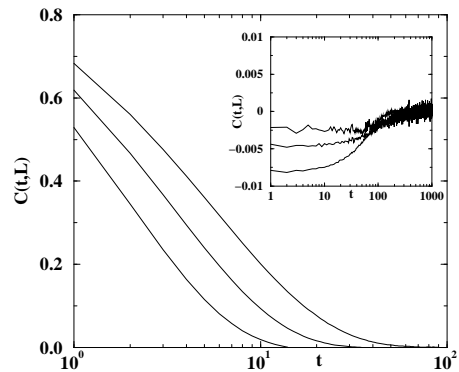}
\end{center}
\caption{
Autocorrelation function of the wave time series
of the undirected and the directed (inset) sandpile 
for $L$=128, 256 and 512.
}
\label{Fig4}
\end{figure}
A more reliable and quantitative check of the validity, or violation, of FSS, is 
based on the evaluation of the various moments of ${\rm Prob}$ \cite{Stella1,Stella2,
Lubeck}. The $q$-th moment is defined as $\langle s^q \rangle = \int s^q {\rm Prob}(s,L)ds$. 
Assuming that FSS holds, it is easy to show that $\langle s^q \rangle \sim L^{\sigma(q)}$
with the moment exponent given by $\sigma(q) = D(q-\tau+1)$ for $q> \tau -1$
and $\sigma(q)=0$ for $0< q <\tau -1$. In the case of multiscaling  $\sigma$ should 
have a nonlinear dependence on $q$. A comparison of $\sigma(q)$ for two given models 
is also a key to establish if they belong to the same universality class, or not.
The value of $\sigma(q)$ was determined 
from the slope of the plot of $\log \langle s^q (L) \rangle$ vs. $\log L$ for $L$ = 1024, 2048 and 4096
with an error $\approx$ 0.01 and for 251 values of $q$ between 0 and 5 (Fig. 3(a)).
The derivative of $\sigma$ is determined by the finite difference method.
Slow but monotonic increase of $d\sigma(q)/dq$ with $q$ clearly indicates the 
multiscaling in BTW as well as in undirected models
(Fig. 3(b)) whereas a saturation of $d\sigma(q)/dq$ indicates validity of FSS for Manna and in directed models (Fig. 3(c)). 
To measure deviations quantitatively
we define a quantity $X_{a,b}=2\frac{|(d\sigma(q)/dq)_a-(d\sigma(q)/dq)_b|}{(d\sigma(q)/dq)_a+(d\sigma(q)/dq)_b}$.
It is observed that after some initial fluctuations $X_{BTW,undir}$ has a maximum value 0.33\% within $q$=2 and 5 whereas
$X_{BTW,direc}$ gradually increases to 5.5 \% at $q$ =5. 
This analysis implies that the undirected model is almost negligibly different
from the BTW model in the $d\sigma(q)/dq$ vs. $q$ plot whereas the deviation of 
directed model from the BTW model is much larger and gradually increases with $q$.
Similarly $X_{Manna,direc}$ is limited witin 0.91 \% whereas $X_{Manna,undir}$
gradually increases to 6.5 \% at $q$ =5 which also implies that the directed model
behaves very similarly to the Manna model, and is very much different from the BTW model.
The above described results concerning $X$ for the various
couples of models are altogether strongly supporting the conclusion that
while the directed model belongs most likely to the Manna universality
class, the undirected one has the multiscaling features known to be
peculiar of the standard BTW sandpile.

By decomposing a large sequence of successive avalanches into waves
in the undirected and directed cases, we obtained
global wave size distributions which obey FSS with the exponents expected for
the BTW model \cite {Prie2} and the Manna model \cite {Stella3}, respectively.
For the BTW sandpile globally sampled waves have a size distribution
with a form as in Eq. (1), with $\tau_w =1$ and $D_w =2$. In the case of the
Manna stochastic sandpile waves can not be defined as in the deterministic
Abelian sandpiles, but a wave-like decomposition was proposed
in Ref.~\cite{Stella3}. The global size distribution of the corresponding
waves obeys FSS with the same exponents obtained for
the avalanche distribution~\cite{Stella3}. As already remarked above
waves can be consistently defined~\cite{Dhar1} in the same way for 
each quenched disorder realization of our directed and undirected models. The global
wave scalings obtained here further support the expectation that
they fall in the Manna and BTW universality classes,
respectively.

Further insight into the different behaviors of the directed
and undirected models can be obtained by analyzing the wave time series
$\{ s_1, s_2, s_3, ..\}$ of the sizes of successive waves as 
in ref. \cite {Stella3}. In Fig. 4 we plot the autocorrelation 
function:

\begin{eqnarray}
C(t,L) = \frac{ \langle s_{k+t} s_k \rangle_L - {\langle s_k \rangle_L}^2 }
{\langle s^2_k \rangle_L - {\langle s_k \rangle_L}^2}
\end{eqnarray}
where the expectation values refer to samples with different $L$ and
include quenched disorder averaging.
The plots 
in Fig. 4 are fully consistent with
similar ones for the BTW and Manna sandpiles \cite {Stella3}.
While in the directed case
the autocorrelation function is essentially zero as soon as $t > 0$,
in the undirected model it grows steadily with $L$, and approximately
scales as $C(t,L) \sim t^{-\tau_c}{\cal G}(t/L^{D_c})$ with 
$\tau_c \approx 0.35$ and $D_c \approx 1$. These exponents should be
compared to 0.40 and 1.02, respectively, as determined for the BTW model \cite {Stella3}.
This long range
autocorrelation must be a consequence of the coherent and uniform
spatial structure of each wave in the undirected case. In the directed model
correlations are destroyed by the much more irregular pattern of topplings, 
with inhomogeneities and holes, in each wave.
The correlation patterns show marked self-averaging,
being reproducible on the basis of very few disorder realizations.

The local out/in degree balance $H_i =H_i'$ at all sites
in the undirected model is essential
for the BTW multiscaling behavior to prevail. Numerically, with quenched disorder
realized as described above, we find that the
density of unbalanced sites with $H_i \ne H_i'$ in the directed model is around 0.88 and those of the
sites with $H_i' >  H_i$ and $H_i' <  H_i$ are equal to 0.44. Now we ask
if there is any critical density of unbalanced sites which demarcates
the behaviors of the undirected and directed models. To study this
we first generated an asymmetric TM in which the fraction of the bonds
with unequal $\Delta$ values is found to be $\approx 0.75$.
We tuned this fraction, and thus the density of unbalanced sites, by randomly selecting these bonds and
making their $\Delta$ values equal by assigning a random integer number between $-1$ and $-m$.
We find that even the presence of as low as $5$ percent bonds with unequal $\Delta$ values is sufficient to
destroy the mutiscaling and to ensure FSS as in the directed model. Thus,
as soon as precise toppling balance is broken, FSS holds, and the universality class turns into that of the 
Manna sandpile. The transition to Manna behavior does not
require a nonzero threshold density of unbalanced sites.
Thus, for the undirected model the symmetry of precise toppling balance is a crucial requisite
for the multiscaling to hold. This requisite is of course satisfied also by the ordinary BTW sandpile.

To conclude, we studied sandpile models with quenched disorder where the elements of 
the TM are randomly assigned. With asymmetric TM       
the precise toppling balance between in- and out-degrees at each site is not maintained.
This imbalance suppresses the wave correlations leading to the BTW-like multiscaling 
behavior of the avalanche size distribution and results a FSS regime in the
universality class of the Manna stochastic sandpile. Thus, a symmetry mechanism
underlies the puzzling difference between BTW and Manna scalings.

We acknowledge discussions with M. De Menech. Work supported by MIUR-COFIN01.

\end{document}